\documentclass[%
 reprint,
 superscriptaddress,
 amsmath,amssymb,
 aps,
 pra,
 nofootinbib
]{revtex4-2}
\usepackage{overpic}
\usepackage{framed}
\usepackage[uline]{hhtensor}
\usepackage{mathrsfs}
\usepackage{mathtools}
\usepackage{xcolor}
\usepackage{textgreek}
\usepackage[caption=false]{subfig}
\usepackage{siunitx}
\usepackage{braket}
\usepackage{amsmath}
\newcommand*{\hbaraux}[2]{\sbox0{\mathsurround=0pt$#1\mkern-1mu\mathchar'26$}\mkern-1mu\lower.07\ht0\box0\mkern-8mu}

\DeclareMathOperator\Log{Log}
\DeclareMathOperator\arctantwo{arctan2}

\newcommand\Eq[1]{Eq.~(\ref{#1})}

\newcommand\rr{\mathbf{r}}

\newcommand\pp{\mathbf{k}}
\newcommand\kk{|\mathbf{k}|}
\newcommand\dk{\mathrm d |\mathbf{k}|}

\newcommand\Flambdark{\mathbf{F}_{\lambda}(\rr,|\pp|)}
\newcommand\Flambdart{\mathbf{F}_{\lambda}(\rr,t)}

\newcommand\Mlambdar{\mathbf{M}_{\lambda}(\rr)}
\newcommand\Mlambdarconj{\mathbf{M}_{\lambda}^\ast(\rr)}

\newcommand\Mlambdafr{\mathbf{M}_{\lambda,f}(\rr)}
\newcommand\Mlambdagr{\mathbf{M}_{\lambda,g}(\rr)}
\newcommand\Mlambdafrconj{\mathbf{M}_{\lambda,f}^\ast(\rr)}
\newcommand\Mlambdagrconj{\mathbf{M}_{\lambda,g}^\ast(\rr)}
\newcommand\Mlambdafrtau{\mathbf{M}_{\lambda,f}^{(\tau_y)}(\rr)}

\newcommand\Frtlambda{\mathbf{F}_{\lambda}(\rr,t)}

\newcommand\Ert{\mathbf{E}(\rr,t)}

\newcommand\Hrt{\mathbf{H}(\rr,t)}

\newcommand\ii{\mathrm{i}}

\newcommand\epsz{\varepsilon_0}

\newcommand\cz{c_0}

\newcommand\op[1]{\mathrm{#1}}

\newcommand\ModeE{\boldsymbol{\mathcal{E}}(\rr)}
\newcommand\ModeEconj{\boldsymbol{\mathcal{E}}^\ast(\rr)}
\newcommand\ModeHconj{\boldsymbol{\mathcal{H}}^\ast(\rr)}
\newcommand\ModeH{\boldsymbol{\mathcal{H}}(\rr)}

\newcommand\ModeEf{\boldsymbol{\mathcal{E}}_f(\rr)}
\newcommand\ModeHf{\boldsymbol{\mathcal{H}}_f(\rr)}

\begin{document}
\title{A scalar product for the radiation of resonant modes}
\author{Maria Paszkiewicz-Idzik}
\thanks{These authors contributed equally.}
\affiliation{Scientific Computing Center, Karlsruhe Institute of Technology, Kaiserstrasse 12, D-76131 Karlsruhe, Germany}
\affiliation{Institute of Theoretical Solid-State Physics, Karlsruhe Institute of Technology, Kaiserstrasse 12, D-76131 Karlsruhe, Germany}
\author{Lukas Rebholz}
\thanks{These authors contributed equally.}
\affiliation{Institute of Theoretical Solid-State Physics, Karlsruhe Institute of Technology, Kaiserstrasse 12, D-76131 Karlsruhe, Germany}
\author{Carsten Rockstuhl}
\affiliation{Institute of Nanotechnology, Karlsruhe Institute of Technology, Kaiserstrasse 12, D-76131 Karlsruhe, Germany}
\affiliation{Institute of Theoretical Solid-State Physics, Karlsruhe Institute of Technology, Kaiserstrasse 12, D-76131 Karlsruhe, Germany}
\author{Ivan Fernandez-Corbaton}
\affiliation{Institute of Nanotechnology, Karlsruhe Institute of Technology, Kaiserstrasse 12, D-76131 Karlsruhe, Germany}
\email{ivan.fernandez-corbaton@kit.edu}
\begin{abstract}
We introduce the conformally-invariant scalar product, originally devised for radiation fields, to the study of the modes of optical resonators. This scalar product allows one to normalize and compare resonant modes using their corresponding radiation fields. Such fields are polychromatic fields free of divergences, which are determined from the complex frequencies and the modal fields on the surface of the resonator. The scalar product is expressed as surface integrals involving the modal fields, multiplied by closed-form factors incorporating the complex frequencies. In a practical application, we study the modes of disk-shaped whispering gallery resonators, and show that the proposed scalar product accurately predicts the geometry-dependent crossings and anti-crossings between modes.
\end{abstract}
\maketitle
\section{Introduction and summary}
Optical resonators allow us to control and enhance light-matter interactions \cite{vahala2003optical}. They offer strong optical confinement, which can reduce the size of optical instruments and decrease optical loss \cite{yuan2022optical}. Resonant re-circulation of an input signal increases the field intensity, which finds applications in lasers and \mbox{photovoltaics \cite{vahala2003optical, yuan2022optical, zhang2020tunable}}. The implementation of optical resonators differs in form, materials, and principle of operation. Among the many kinds of optical resonators, the ones hosting whispering gallery modes (WGMs) are particularly attractive for many applications due to their high quality factor and unique spectral properties, such as tunability, narrow linewidth, and high stability \cite{ilchenko2006optical, yuan2022optical,loyez2023whispering}. 

The physics of resonators can be conveniently studied through their resonant modes \cite{Kristensen_Modeling_2020}, which are the natural damped resonances of the system. Such modes are also known as quasi-normal modes, leaky modes, electromagnetic eigenmodes, or simply, modes. In particular, resonant modes are being used for the study and engineering of light-matter interactions \cite{Lalanne2019, sauvan2022normalization,Ge2015,Kamandar2018,Binkowski2020,Colom2018,Wu2021,ge2014quasinormal,Gorkunov2024, kristensen2012generalized,franke2019quantization,gippius2010resonant, coenen2013resonant,martin2019chiral,Wu2024,betz_efficient_2024}. For example, the fields scattered by the resonator upon a given illumination may be expanded to good approximation as a linear combination of a few modal fields, at least in limited frequency ranges \cite{ren2022connecting,kamandar2017quantum}. The general question of orthogonality and completeness of the modal fields outside the resonator is particularly relevant for such applications \cite{both_resonant_2021}. Such a question is complicated by the divergence of time-harmonic modal fields as $|\rr| \rightarrow \infty$, albeit, in principle, such divergence can be mitigated by causality \cite{Abdelrahman2018,Colom2018}. In this context, a crucial question is {\em which scalar product to use for normalization and projections?} These questions have received considerable attention in recent times. Reference~\cite{sauvan2022normalization} contains a comprehensive review of the properties and suitability of several different approaches \cite{kristensen2012generalized,Muljarov:18,Lalanne2018}.  

For radiated fields, there is a scalar product with many desirable properties \cite{Gross1964}. For example, the square of the norm induced by such scalar product $\braket{f|f}$, gives the number of photons of the field \cite{Zeldovich1965}. Also, the values of fundamental quantities in the field, such as energy or momentum, can be computed as ``sandwiches'' $\braket{f|\Gamma|f}$, where $\Gamma$ is the operator representing the particular fundamental quantity \cite[Chap.~3,\S 9]{Birula1975}. For example, using the energy operator $\mathrm{H}$, the value of $\braket{f|\mathrm{H}|f}$ can be shown to be identical to the typical integral giving the energy of the field. Such scalar product has the unique property of being invariant upon any conformal transformation \cite{Gross1964}. That is, the scalar product between transformed fields is equal to the scalar product between the original fields. Since the conformal group is the largest group of invariance of Maxwell's equations \cite{Bateman1910}, the conformal invariance of the scalar product is a very strong argument in favor of its use. For example, it is precisely those invariance properties that underpin the consistent frame-independent definition of projective measurements in electromagnetism \cite[Sec.~III]{FerCor2022b}. The use of this scalar product in light-matter interactions has been recently reviewed \cite{FerCor2024b}. 

Here, we introduce the use of the conformally-invariant scalar product for radiation fields in the study of optical resonators. We start by deriving a polychromatic radiation field free of divergences from the field profile and the complex frequency of a given electromagnetic eigenmode of a three-dimensional (3D) structure. Then, a cross-energy expression between the radiations of any two given eigenmodes, $\braket{f|\mathrm{H}|g}$, is identified as a suitable scalar product. The computation of $\braket{f|\mathrm{H}|g}$ consists of integrals of simple functions of the modal fields at the surface of the resonator, and closed-form factors involving the complex modal frequencies (see \Eq{eq:esp}). In contrast with several existing expressions, \Eq{eq:esp} does not involve the material parameters of the resonator, or the exponentially growing fields with complex wavenumber. Boundary element methods such as those described in \cite{Hohenester2012,Hohenester2022} are perfectly suited for the computation of the cross-energy scalar product.

We show that the cross-energy scalar product produces physically consistent results, by means of a first exemplary application. We study the modes of a WGM disk resonator as the thickness of the disk changes \cite{woska2022intrinsic}. We observe that the absolute value of $\braket{f|\mathrm{H}|g}$ between two normalized modes is zero in particular when the real frequencies of that particular pair of modes cross. In sharp contrast, for anti-crossing modes, the same quantity shows a prominent peak that grows as the real frequencies of that particular pair of modes get close to each other. The position of the peak coincides with the thickness for which the two modes anti-cross, and for which the modal overlap inside the resonator is maximized. The increasing non-orthogonality implies a decreasing distinguishability of the modes in the far field. 

The rest of the article is organized as follows. In Section~\ref{sec:esp}, we obtain the field radiated by a given leaky mode by using the Mittag-Leffler theorem. The resulting polychromatic field is free of divergences. We insert these fields into a recent \cite{vavilin2024electromagnetic} surface integral expression for $\braket{f|\mathrm{H}|g}$, and derive the corresponding expression for radiation fields of resonant modes of 3D structures, which we call the cross-energy scalar product. In Section~\ref{sec:wgm}, we apply the cross-energy scalar product to the resonant modes of a disk-shaped WGM resonator, with the particular aim of studying the case of pairs of  modes exhibiting avoided crossings as the thickness of the disk changes. We finish with conclusions and an outlook in Section~\ref{sec:con}.

\section{A scalar product for the radiation of leaky modes from 3D structures\label{sec:esp}}
We start by obtaining the field radiated by a given leaky mode. The resulting polychromatic field is free of divergences.

\subsection{Electromagnetic fields radiated by a leaky mode}
\begin{figure}[h!]
	\includegraphics[width=0.8\linewidth,trim=150 130 150 110, clip]{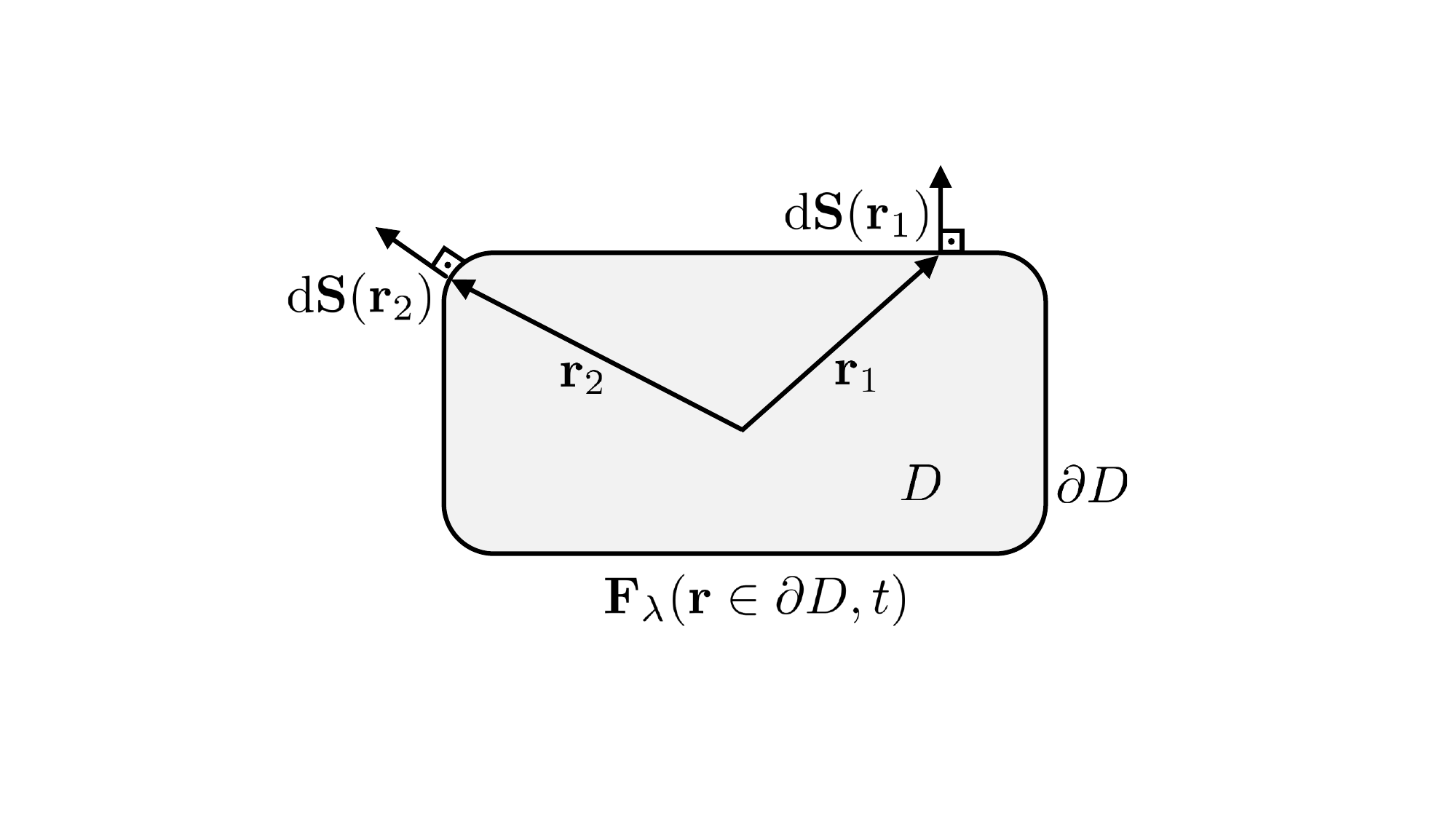}
	\caption{A volume $D$ in $\mathbb{R}^3$ is delimited by a closed surface $\partial D$ with continuous first derivative. Helical fields ${\mathbf{F}_{\lambda=\pm 1}(\rr\in \partial D,t)}$ on the surface produce electromagnetic radiation towards the outside of $D$. The d$\mathbf S(\rr)$ are outwards-pointing normal vectors of the surface element at each point $\rr\in\partial D$. \label{fig:DD} } 
\end{figure}

Let us consider Fig.~\ref{fig:DD}, where a closed surface \mbox{in $\mathbb{R}^3$} delimits a volume $D$ with a boundary $\partial D$ that has continuous first derivatives. It is surrounded by an achiral, non-absorbing, homogeneous, and isotropic background medium, which we assume to be vacuum for simplicity,  but without loss of generality. Any other such surrounding medium can be readily accommodated in the formulas by replacing the vacuum permittivity and permeability by those of the medium. We assume the existence of time-dependent helical fields $\mathbf{F}_{\lambda}(\rr\in \partial D,t)$, for helicity $\lambda=\pm1$, on the boundary surface $\partial D$. We will later work with their monochromatic components $\Flambdark$. For fields that contain only positive frequencies, the helical fields are defined as:
\begin{equation}
	\label{eq:lambda}
	\Frtlambda=\sqrt{\frac{\varepsilon_0}{2}}\left[\Ert+\ii\lambda Z_0\Hrt\right]\,,
\end{equation}
with vacuum permittivity $\varepsilon_0$, vacuum impedance $Z_0$, time and spatially dependent complex electric field $\Ert$, and complex magnetic field $\Hrt$. The helical fields split any electromagnetic field into its left and right circular polarization handedness, with $\lambda=1$ and $\lambda=-1$, respectively.

An expression for the scalar product between radiation fields that only involves integrals of the fields over closed spatial surfaces was recently derived \cite{vavilin2024electromagnetic}. In particular, the $\mathbf{F}_{\lambda}(\rr\in \partial D,|\pp|)$ appear in the expressions for the number of photons $\braket{f|f}$, and energy $\braket{f|\mathrm{H}|f}$ of a given field $\ket{f}$, which can be computed as integrals on a closed boundary \cite[Eqs.~(19, 21)]{vavilin2024electromagnetic}:
\begin{multline}\label{eq:nphotons}
    \braket{f|f} = \sum_{\lambda=\pm1}(-\ii\lambda) \int_{>0}^\infty \frac{\mathrm d\kk}{\hbar c_0 \kk}\\
	\int_{\rr\in\partial D}\mathrm d\mathbf S(\rr)\cdot [\mathbf F^*_\lambda (\rr,\kk)\times \mathbf F_\lambda (\rr,|\mathbf k|)]\,,
\end{multline}
\begin{multline}\label{eq:energy_F}
	\braket{f|\mathrm {H}|f} = \sum_{\lambda=\pm1}(-\ii\lambda) \int_{>0}^\infty \dk \\
    \int_{\rr\in\partial D}\mathrm d\mathbf S(\rr)\cdot [\mathbf F^*_\lambda (\rr,\kk)\times \mathbf F_\lambda (\rr,|\mathbf k|)]\,,
\end{multline}
where $\cz$ is the speed of light in vacuum, $\hbar$ is Planck's constant divided by $2\pi$, $\mathrm H$ the energy operator, $\mathrm d\mathbf S$ the infinitesimal surface element at the position $\rr$, and the integral is over any piecewise smooth surface $\partial D$ enclosing a compact volume containing the sources of radiation.

The fact that electromagnetic fields on a boundary act as sources of radiation fields is well-known in electromagnetism (see, {\it e.g.}, \cite{Kress2001}\cite[Chap.~5]{Lakhtakia1994}). The components of the electric and magnetic fields tangential to the surface uniquely determine the whole field outside the enclosed volume. This enables us to obtain the field radiated by a given leaky mode. It is important to note that the boundary $\partial D$ does not need to be twice continuously differentiable as stated in, {\it e.g.}, \cite{Kress2001}, rather once continuously differentiable is enough \cite[Chap.~5]{Lakhtakia1994} because Maxwell's equations for helical fields only contain first-order derivatives \cite[\S 1-2.3]{Lakhtakia1994}. This condition increases the class of surfaces to which the results apply from ``continuous curvature'' surfaces such as an ellipsoid, to ``continuous tangent'' surfaces such as a disk with rounded corners.

Let us now assume that the boundary $\partial D$ in Fig.~\ref{fig:DD} is the boundary of an object, and that the electromagnetic eigenmodes of the object are available to us. Such eigenmodes will, in general, be leaky, and each of those modes with finite lifetimes can be characterized by spatial field profiles $\left\{\ModeE,\ModeH\right\}$, and complex frequencies $\omega = \Omega-\ii \Gamma$, with $\{\Omega, \Gamma\} \in \mathbb R^+$. Eigenmodes come in conjugate pairs \cite[Sec.~2.3]{Lalanne2018}: If there exists a leaky mode with complex frequency $\omega=\Omega-\ii \Gamma$, and modal fields $\ModeE$ and $\ModeH$, then there exists another leaky mode with complex frequency $-\omega^*$ and modal fields $\ModeEconj$ and $\ModeHconj$. From now on, and unless otherwise specified, we assume that $\rr$ is a point on the surface of the object.

We want to obtain the $\Flambdark$ that specify a physical radiation field. Importantly, the $\Flambdark$ must meet the condition
\begin{equation}
	\label{eq:mconj}
	\mathbf{F}_\lambda(\rr,|\pp|)=\mathbf{F}^*_\lambda(\rr,-|\pp|),
\end{equation}
because in electromagnetism the same information must be contained in the two sides of the real frequency axis \cite[\S 3.1]{Birula1996}. After a two-sided Fourier transform, the condition in \Eq{eq:mconj} ensures $(\rr,t)$-dependent real-valued fields.

We advance towards the specification of the $\Flambdark$ for leaky modes by considering the complex frequencies $\omega$ and $-\omega^*$ as the two poles of a complex vectorial function $\mathbf{F}_\lambda(\rr,z\in\mathbb{C})$, whose respective residues are the helical combinations of the electric and magnetic modal fields
\begin{equation}
	\label{eq:resonant_state}
	\begin{split}
		\omega=\Omega-\ii \Gamma \,,\quad& \Mlambdar=\sqrt{\frac{\epsz}{2}}\left[\ModeE+\ii\lambda Z_0\ModeH\right] \,,\\
		-\omega^\ast=-\Omega-\ii \Gamma\, ,\quad& \Mlambdarconj=\sqrt{\frac{\epsz}{2}}\left[\ModeE+\ii\lambda Z_0\ModeH\right]^\ast \,.
	\end{split}
\end{equation}
At first sight, and in light of \Eq{eq:lambda}, $\Mlambdarconj=\sqrt{\frac{\epsz}{2}}\left[\ModeEconj-\ii\lambda Z_0\ModeHconj\right]$ seems to be a field with helicity $-\lambda$, but this is not so: such field is also a field with helicity $\lambda$ because the definition of handedness changes sign with the sign of the real part of the frequency \cite[p.~210]{Kaiser2010},\cite[Sec.~3.1]{Birula1996}.

The sought after $\Flambdark$ can be obtained using the Mittag-Leffler theorem \cite[p.~515]{Arfken2012}, which allows one to expand a function $f(z)$ of a complex variable $z$ as: 
\begin{equation}
	f(z)=f(0)+\sum_{n=1}^\infty b_n\left(\frac{1}{z-z_n}+\frac{1}{z_n}\right).
\end{equation}

For the expansion to hold, $f(z)$ should be analytical everywhere, excluding infinity and a set of discrete points $z_n$ with residues $b_n$, and $|f(z)/z|\rightarrow 0$ as $|z|\rightarrow\infty$. We assume that each scalar component of the vectors $\mathbf{F}_\lambda(\rr,z)$ meets such conditions. This kind of assumption underlies the common use of the Mittag-Leffer theorem in the context of resonant modes \cite[Sec.~2.3]{sauvan2022normalization}. One can then write:
\begin{equation}
    \label{eq:F_mittag-leffler}
	\begin{split}
		&\mathbf{F}_\lambda(\rr,k)=\mathbf{F}_\lambda(\rr,0)+\\
		&\frac{\ii\cz}{\sqrt{2\pi}}\left[\frac{\Mlambdar}{\cz k-\omega}+\frac{\Mlambdar}{\omega}+\frac{\Mlambdarconj}{\cz k + \omega^*}-\frac{\Mlambdarconj}{\omega^*}\right],
	\end{split}
\end{equation}
where $k\in\mathbb{R}$. The expression in \Eq{eq:F_mittag-leffler} meets the requirement in \Eq{eq:mconj}.

For $\mathbf{F}_\lambda(\rr,k)$ to be suitable for the inverse Fourier transform, it must vanish as $k\to \pm\infty$. We are hence led to set 
\begin{equation}
	\label{eq:zero}
	\mathbf{F}_\lambda(\rr,0)=-\frac{\ii\cz}{\sqrt{2\pi}}\left[\frac{\Mlambdar}{\omega}-\frac{\Mlambdarconj}{\omega^*}\right].
\end{equation}
Finally, since we are interested in $k=|\pp|>0$, we reach:
\begin{equation}
	\label{eq:k}
\Flambdark=\frac{\ii c_0\Mlambdar}{\sqrt{2\pi}\left(c_0|\pp|- \omega\right)}+\frac{\ii c_0\Mlambdarconj}{\sqrt{2\pi}\left(c_0|\pp|+ \omega^*\right)}.
\end{equation}

Real-valued time-domain fields at each point $\rr$ of the surface of the object can be obtained with the inverse Fourier transform,
\begin{equation}
	\label{eq:czp}
	\begin{split}
    	&\Flambdart=\int_{-\infty}^\infty \frac{\text{d} k}{\sqrt{2\pi}}\ \mathbf{F}_\lambda(\rr,k)\exp(-\ii c_0 k t)\\
		&=2\mathrm{Re}\left\{
            \int_{0}^\infty \frac{\dk}{\sqrt{2\pi}}\ 
            \Flambdark \exp(-\ii c_0\kk t)
        \right\},
	\end{split}
\end{equation}
which results in:
\begin{equation}
	\label{eq:rt}
	\begin{split}
		&\mathbf{F}_\lambda(\rr,t)=\\
		&\left[\Mlambdar\exp\left(-\ii\Omega t\right)+\Mlambdarconj\exp\left(\ii\Omega t\right)\right]\exp(-\Gamma t)u(t),
	\end{split}
\end{equation}
where $u(t)$ is the Heaviside step function.

The Heaviside function avoids the amplification that occurs in $\exp\left(-\Gamma t\right)$ for $t<0$ while keeping the damping that occurs for $t>0$. Moreover, and importantly, the fields $\Flambdark \exp\left(-\ii c_0|\pp|t\right)$ result in radiated fields that decay as $1/|\rr|$ outside the resonator when $|\rr|\rightarrow\infty$ \cite[Eq.~(35)]{Kress2001}, also avoiding the divergence of the modal fields as $|\rr|$ grows. Hence, the presented strategy avoids the two exponential growths of the modal field outside the resonator, as $t\rightarrow -\infty$, and as $|\rr|\rightarrow\infty$, which have been tied to each other by causality \cite{Abdelrahman2018}. Here, the wavenumbers of the radiation fields are always real. In contrast, taking the complex eigenfrequency to imply a complex wavenumber causes the field outside the resonator to grow exponentially as $|\rr|$ increases. While such growing fields are being used for expanding scattered fields, typically in the monochromatic case, the exponential growth is clearly incompatible with a physical emission.

The appearance of $u(t)$ has the following physical interpretation. The surface fields specified in \Eq{eq:k} or \Eq{eq:rt} will radiate an outgoing polychromatic pulse. Such pulse is the radiative decay of energy stored inside the object by that given conjugate pair of leaky modes. The time $t=0$ is the start of the emission. The procedure that lead us to \Eq{eq:k} does not make any assumption regarding how the energy was previously acquired by the leaky modes. From now on, when we refer to a leaky mode, we mean one such conjugate pair.

We note that, even though the surface fields that act as sources must be defined on the surrounding medium, and $\left\{\ModeE,\ModeH,\ModeEconj,\ModeHconj\right\}$ are the fields inside the object, this is not a problem because the tangential components of the electric and magnetic fields are continuous at the interface between the object and the surrounding medium. The surface integrals in \Eq{eq:nphotons} and \Eq{eq:energy_F} ensure that only the components tangential to the surface affect the result. In this way, the potentially complicated material functions characterizing the object, and their derivatives, will not appear in the expression for the scalar product that we propose in the next section, in contrast with many of the existing expressions. 

There are many computational strategies for obtaining the $\left\{\Mlambdar,\;\omega\right\}$ \cite[Sec.~3]{Lalanne2018}. For example, for sufficiently small 3D objects the leaky modes can be obtained with use of software packages such as JCMsuite \cite{JCM}, Lumerical \cite{lumerical}, or COMSOL \cite{comsol}, to name just a few.

\subsection{The cross-energy scalar product}
Equation~(\ref{eq:nphotons}) can be used to compute the number of photons of a leaky mode by inserting the monochromatic components from \Eq{eq:k}. However, there are difficulties in evaluating the corresponding expression for a scalar product between two leaky modes, $\braket{f|g}$. If we instead focus on the expression for the radiated energy, we find that the cross-energy $\braket{f|\mathrm H|g}$, between arbitrary modes $f$ and $g$, meets the requirements of a scalar product, which can be readily verified using the fact that the energy operator $\mathrm H$ is self-adjoint. The use of a cross-energy scalar product is reminiscent of the scalar product derived from the Poynting vector for the propagating modes of waveguides \cite{hiremath2005analytic, Hiremath:2006}, which can be used to investigate the orthogonality between the modes \cite{Paszkiewicz:24}. This motivates us to use $\braket{f|\mathrm H|g}$ as the scalar product for leaky modes. It is obtained by inserting \Eq{eq:k} into \Eq{eq:energy_F}, and we call it the cross-energy scalar product: 
\begin{widetext}
\begin{equation}\label{eq:espA}
\begin{split}
   \braket{f|\mathrm H|g} 
   &= 
   \sum_{\lambda=\pm1}(-\ii\lambda) 
   \int_{>0}^\infty \dk 
   \int_{\rr\in\partial D}\mathrm d\mathbf S(\rr)
   \cdot [
        \mathbf F_\lambda (\rr,\kk)^\ast
        \times \mathbf G_\lambda (\rr,|\mathbf k|)
    ] 
    \\
    &= 
    \sum_{\lambda=\pm1}(-\ii\lambda)
    \frac{c_0^2}{2\pi}\Bigg\{ 
        \int_{>0}^\infty\dk
        \frac{1}{\left(c_0|\pp| - \omega_f^\ast\right)}
        \,\frac{1}{\left(c_0|\pp| - \omega_g\right)} 
        \int_{\rr\in\partial D}\mathrm d\mathbf S(\rr)\cdot\left[ 
            \Mlambdafrconj \times \Mlambdagr
        \right]
        \\
        &\qquad\qquad\qquad\quad
        + \int_{>0}^\infty\dk
        \frac{1}{\left(c_0|\pp| - \omega_f^\ast\right)}
        \,\frac{1}{\left(c_0|\pp| + \omega_g^\ast\right)} 
        \int_{\rr\in\partial D}\mathrm d\mathbf S(\rr)\cdot\left[ 
            \Mlambdafrconj \times \Mlambdagrconj
        \right]
        \\
        &\qquad\qquad\qquad\quad
        + \int_{>0}^\infty\dk
        \frac{1}{\left(c_0|\pp| + \omega_f\right)}
        \,\frac{1}{\left(c_0|\pp| - \omega_g\right)} 
        \int_{\rr\in\partial D}\mathrm d\mathbf S(\rr)\cdot\left[ 
            \Mlambdafr \times \Mlambdagr
        \right]
        \\
        &\qquad\qquad\qquad\quad
        + \int_{>0}^\infty\dk
        \frac{1}{\left(c_0|\pp| + \omega_f\right)}
        \,\frac{1}{\left(c_0|\pp| + \omega_g^\ast\right)} 
        \int_{\rr\in\partial D}\mathrm d\mathbf S(\rr)\cdot\left[ 
            \Mlambdafr \times \Mlambdagrconj
        \right]
    \Bigg\}
    \,,
\end{split}
\end{equation}
where the subscripts $f$, $g$ denote the two considered modes. Appendix~\ref{appendixA} shows that the result of the $\dk$ integral can be obtained in closed form, with which one obtains:
\begin{equation}\label{eq:esp}
\boxed{
\begin{aligned}
    \braket{f|\mathrm H|g} 
    &=
    \frac{c_0}{2\pi}
    \sum_{\lambda=\pm1}(-\ii\lambda)
        \frac{\Log(-\omega_f^\ast) - \Log(-\omega_g)}{-\omega_f^\ast + \omega_g}
        \int_{\rr\in\partial D}\mathrm d\mathbf S(\rr)\cdot\left[ 
            \Mlambdafrconj \times \Mlambdagr
        \right]
        \\
        &+ \frac{c_0}{2\pi}
        \sum_{\lambda=\pm1}(-\ii\lambda)
        \frac{\Log(-\omega_f^\ast) - \Log(\omega_g^\ast)}{-\omega_f^\ast - \omega_g^\ast}
        \int_{\rr\in\partial D}\mathrm d\mathbf S(\rr)\cdot\left[ 
            \Mlambdafrconj \times \Mlambdagrconj
        \right]
        \\
        &+ \frac{c_0}{2\pi}
        \sum_{\lambda=\pm1}(-\ii\lambda)
        \frac{\Log(\omega_f) - \Log(-\omega_g)}{\omega_f + \omega_g}
        \int_{\rr\in\partial D}\mathrm d\mathbf S(\rr)\cdot\left[ 
            \Mlambdafr \times \Mlambdagr
        \right]
        \\
        &+ \frac{c_0}{2\pi}
        \sum_{\lambda=\pm1}(-\ii\lambda)
        \frac{\Log(\omega_f) - \Log(\omega_g^\ast)}{\omega_f - \omega_g^\ast}
        \int_{\rr\in\partial D}\mathrm d\mathbf S(\rr)\cdot\left[ 
            \Mlambdafr \times \Mlambdagrconj
        \right]
    \,.
\end{aligned}
}    
\end{equation}
\end{widetext}
In the following section, we use the cross-energy scalar product to study the modes of \mbox{a disk} resonator as the thickness of the disk changes.

\section{Application to whispering gallery resonators\label{sec:wgm}}
\subsection{The modes of a disk}
\begin{figure}[h!]
	\includegraphics[width=\linewidth,trim=0 40 0 80, clip]{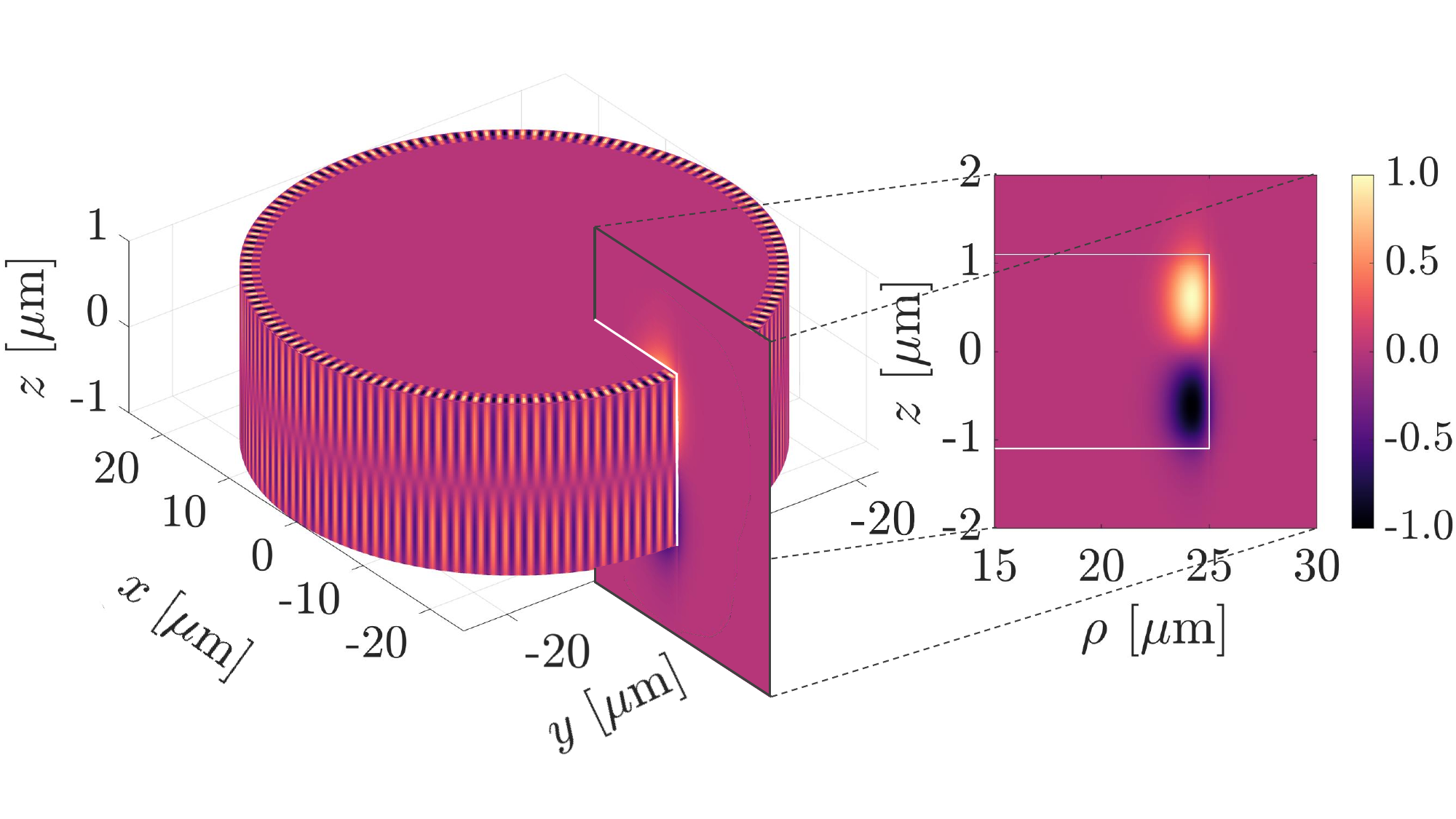}
	\caption{\label{fig:disk_projected} The radial component of the real part of the electric field of the fundamental mode on the disk surface (left). Because of the azimuthal symmetry, we consider resonant modes computed in a slice of the disk in the $xz$ plane (projected on the right). The discretization in the radial direction is nonuniform, and finer close to the rim. The field decreases by several orders of magnitude at the borders of the chosen computational domain. For the sake of visibility, the domain in the figure differs from the size of the actual computational domain, and the scale of the $z$-axis is different from that of the other two axes.}
\end{figure}

Here, we consider a WGM resonator, modeled as \mbox{a disk} of radius \SI{25}{\micro\meter} and thickness changing from \SI{1.8}{\micro\meter} to \SI{3.1}{\micro\meter}. The electric and magnetic fields and the resonance frequencies are computed with the finite element method (FEM) simulation software \mbox{JCMsuite}. It solves the resonant mode problem by finding electric and magnetic fields ($\ModeEf$, $\ModeHf$) of the modes $f$ and corresponding eigenvalues \mbox{$\omega_f=\Omega_f-\ii \Gamma_f$}, satisfying the time-harmonic Maxwell equations with outgoing boundary conditions in \mbox{a source-free} medium. In our case, the numerical implementation exploits the rotational symmetry of the disk with respect to the $z$-axis, and the eigenmode computation is reduced to a two-dimensional problem in the plane ($\rho=\sqrt{x^2+y^2}$, $\varphi=\arctantwo(y,x)=0$, $z$), as visualized in Fig.~\ref{fig:disk_projected}. At any point of the disk, the fields of a given solution (in Cartesian coordinates) obey
\begin{equation}\label{eq:M_JCM}
	\boldsymbol{\mathcal{E}}^{(m^f)}_f(\rho\cos\varphi, \rho\sin\varphi, z) = \op{R} \cdot \boldsymbol{\mathcal{E}}^{(m^f)}_f(\rho,0,z)\mathrm e^{\ii m^f\varphi}\,,
\end{equation}
and similarly for $\boldsymbol{\mathcal{H}}^{(m^f)}_f$, with azimuth $\varphi$, integer azimuthal mode number $m^f$, and the rotation matrix\footnote{Note, that JCMsuite uses a different convention where the $y$-axis is the axis of rotational symmetry.} \cite{JCM}
\begin{equation}\label{eq:R}
\op{R} = \left[\begin{array}{ccc}
    \cos\varphi & -\sin\varphi & 0 \\
    \sin\varphi & \cos\varphi & 0 \\
    0 & 0 & 1
\end{array}\right]\, .
\end{equation}

The fields ($\boldsymbol{\mathcal{E}}^{(m^f)}_f$, $\boldsymbol{\mathcal{H}}^{(m^f)}_f$) can be used to construct (helical) modal fields $\mathbf{M}_{\lambda,f}^{(m^f)}(\rr)$ following the definition in \Eq{eq:resonant_state}, and then then corresponding $\mathbf{F}_{\lambda}^{m^f}(\rr,|\pp|)$ in \Eq{eq:k}. It is straightforward to show that under the cross-energy scalar product two modes are orthogonal unless ${\left|m^f\right|=\left|m^g\right|}$: For instance, the surface integral in the first line of \Eq{eq:esp} contains a term $\mathrm e^{\ii (m^g-m^f)\varphi}$, which makes the integral vanish identically if ${m^g \neq m^f}$. Importantly, modes of opposite azimuthal mode number, that is, for ${m^f = - m^g}$, can be non-orthogonal. This is because the monochromatic components in \Eq{eq:k} contain the modal fields and their complex conjugate and, therefore, combine terms of azimuthal dependence $\mathrm{e}^{\pm\ii m\varphi}$. This can be understood as a direct consequence of \Eq{eq:mconj}, which is a necessary condition for real-valued fields in $(\rr, t)$. 

This non-orthogonality motivates one to find a different way to build modes for the analysis of our resonator. To such end, we consider a mirror reflection across a plane that contains the symmetry axis of the disk, for example $\op{M}_y$, which acts by changing the sign of the y-coordinate. Such operation is a symmetry of the disk, which means that applying it to a given leaky mode produces a potentially different mode, but with the same modal frequency. For cylindrically symmetric systems that are mirror symmetric with respect to a plane that includes the axis of rotation, such as the disk, the corresponding modal field $\mathbf{M}_{\lambda,f}^{(-m^f)}(\rr)$ is readily found by the mirror reflection, which transforms the fields in \Eq{eq:M_JCM} into those that exhibit an azimuthal dependence of $\mathrm e^{\ii (-m^f)\varphi}$. See Appendix~\ref{appendixB} for details. The fields obtained in this way for the disk resonator match the ones numerically computed for $-m^f$ precisely up to phase factors, which are explained by the rotational degree of freedom in choosing a mirror plane that contains the symmetry axis. It therefore suffices to compute the fields only for one sign of $m$.

We show in Appendix~\ref{appendixB} that the leaky modes constructed with \Eq{eq:k}, using the following even and odd (${\tau_y=\pm1}$) combinations
\begin{equation}\label{eq:even-odd}
    \Mlambdafrtau = \frac{1}{\sqrt{2}}\left(\mathbf{M}_{\lambda,f}^{(m^f)}(\rr) + \tau_y \mathbf{M}_{\lambda,f}^{(-m^f)}(\rr)\right) \,,
\end{equation}
are orthogonal when their $\tau_y$ values are not the same, independently of the values of their azimuthal numbers. Note that the mirror transformation relates the fields with opposite signs of $\lambda$, since it changes the handedness of the fields. The combinations in \Eq{eq:even-odd} result in leaky modes that are eigenstates of $\op{M}_y$ with eigenvalue $\tau_y$. They are also eigenstates of rotations along the $z$-axis by discrete angles $\pi/|m^f|$. We will work with the $\op{M}_y$-symmetric modes constructed with \Eq{eq:even-odd}, defined for $\tau_y=\pm1$ and, without loss of generality, with $m^f\ge 0$.

Let us now advance to the study of the modes of the disk resonator, selecting ${|m|=139}$ to match \cite{woska2022intrinsic}. The modes are calculated for a real permittivity \mbox{$\varepsilon_\text{res}^\text{r}=(1.481)^2$} and imaginary permittivity $\varepsilon_\text{res}^\text{i}=10^{-4}$ of the disk resonator, and the surrounding permittivity of air $\varepsilon_\text{sur}=(1.000275)^2$, following \cite{woska2022intrinsic}. Note that the surrounding medium here is not vacuum as assumed in the construction of the fields before. However, $\varepsilon_\text{sur}$ differs only marginally from unity and we can neglect its influence, since the results are virtually identical to the vacuum case. Due to the field localization near the resonator rim, the region from which the fields are extracted can be limited to $[19.5, 27.0]$ \SI{}{\micro\meter} along $\rho$ and $[-3,3]$ \SI{}{\micro\meter} in $z$-direction. The target relative precision of resonance frequencies is set to $10^{-6}$, and an adaptive mesh refinement scheme with two maximum refinement steps is used. The computational domain is surrounded by perfectly matched layers (PMLs) \cite{JCM}.

\begin{figure}[h!]
	\includegraphics[width=\linewidth,trim=0 110 0 110, clip]{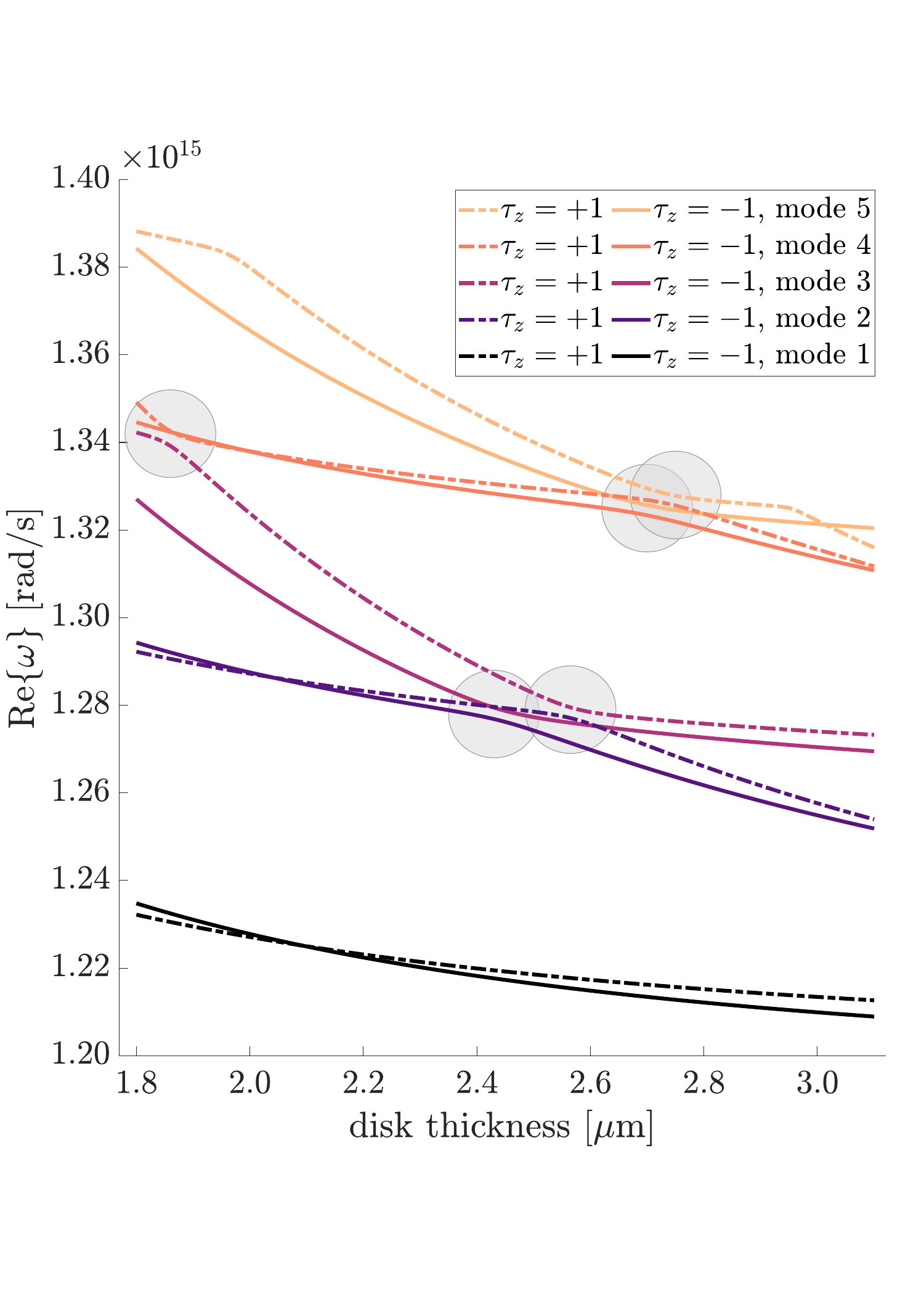}
	\caption{\label{fig:anticrossings} Real part of the angular resonance frequency of the five pairs of lowest order modes as a function of disk thickness. Solid lines follow modes with negative $\tau_z$, and dashed lines mark the resonance frequencies of modes with positive $\tau_z$, where $\tau_z$ is the eigenvalue of the mode upon a $z\mapsto -z$ mirror transformation $\op{M}_z$. Modes of equal $\tau_z$ undergo avoided crossings at the spots marked with gray circles.}
\end{figure}

The resonance frequencies of the modes change with varying thickness of the disk. The real parts of the frequencies of the first ten modes emerging from the FEM computation are plotted in Fig.~\ref{fig:anticrossings} for disk thicknesses ranging from \SI{1.8}{\micro\meter} to \SI{3.1}{\micro\meter}. As a result of the FEM computation, the modes are sorted independently for every disk thickness, and assigned \mbox{a spectral} order number according to their increasing real part of the resonance. However, it is known from symmetry analysis in \cite{woska2022intrinsic} that some modes exhibit crossings over the course of the varying disk thickness, and some exhibit avoided crossings, also called anti-crossings. In particular, the disk is invariant under the mirror reflection $\op{M}_z$: $z \mapsto -z$. Then, the electric field distribution $\ModeEf$ of each mode is an eigenstate of such reflection with eigenvalue $\tau_z=1$ or $\tau_z=-1$, and the magnetic field distribution $\ModeHf$ is also an eigenstate of the reflection $z \mapsto -z$ with the opposite eigenvalue $-\tau_z$. The sign difference is due to the polar and axial character of electric and magnetic fields, respectively. For example, the field in Fig.~\ref{fig:disk_projected} transforms with an eigenvalue $\tau_z=-1$. Modes of opposite $\tau_z$ cross and modes with the same $\tau_z$ anti-cross, as seen in \cite{woska2022intrinsic}. The anti-crossings are marked with gray circles in Fig.~\ref{fig:anticrossings}. 

It is important to highlight that crossings affect the ordering of the modes. The spectral order coming from the numerical tool naturally swaps the labels for modes that cross as the thickness of the disk increases, and, therefore, does not reflect their true order. In our case, however, this can be circumvented by considering the spectral order of modes separately for different $\tau_z$, since crossings only occur between modes of opposite $\tau_z$. More complex resonators may not allow such ``manual'' tracking. Then, the cross-energy scalar product between eigenmodes can be used as a general way to track the modes as some resonator parameters change smoothly. This can be done by projecting the radiation field of each mode for a given set of parameters onto the radiation field of each mode in the next set, and connecting a given mode of the first set to the mode of the second set which results in the maximum value of their mutual projection. 

\subsection{The cross-energy scalar product. Similarity between modes}

We analyze the orthogonality properties of the modes as a function of the disk thickness using the following quantity:
\begin{equation}\label{eq:normalized_scalar_product}
    |\braket{\hat f|\mathrm H|\hat g}|^2 =  \frac{\left|\braket{f|\mathrm H|g}\right|^2}{\braket{f|\mathrm H|f}\braket{g|\mathrm H|g}}\,, 
\end{equation}
where the normalization of each mode is considered so that each normalized mode radiates the same amount of energy: $\ket{\hat f}=\ket{f}/\sqrt{\braket{f|\mathrm H|f}}$, and $\ket{\hat g}=\ket{g}/\sqrt{\braket{g|\mathrm H|g}}$.

The expression in \Eq{eq:esp} is further simplified by exploiting the cylindrical symmetry of the system, which reduces the two-dimensional surface integrals to one-dimensional contour integrals. See Appendix~\ref{appendixB} for details. The integral is then computed using the trapezoidal rule over the contour $\mathcal C$ drawn with a white line in the projection in Fig.~\ref{fig:disk_projected}. The discretization of the contour has been chosen with enough resolution so that the simple trapezoidal rule produces satisfactory results, as shown by convergence analysis. Higher-order quadrature rules can be beneficial for reducing the number of sampling points for a given desired accuracy, in particular for asymmetric resonators where the integrals must be carried out over the whole surface of the resonator \cite{Hohenester2012,Hohenester2022}. We choose the dimensions of the contour (the height and width) to be 2 \% larger than the actual contour of the $\varphi=0$ slice of the disk to avoid regions of numerical artifacts in the simulated fields, which are found near the edges of the disk resonator. Sampling the fields slightly outside the resonator technically means sampling fields that feature an exponentially divergent behavior towards spatial infinity due to being associated with a complex wavenumber. However, the rate of divergence is related to the radiative damping of the resonant modes. For fairly high-quality resonances, indicated by the ratio $\Omega / \Gamma$, in the order of $10^{5}$ for the modes discussed here, the exponential divergence is slow enough to be negligible this close to the resonator surface. Correspondingly, this is a good approximation to the desired fields at the surface of the disk. The contour is discretized finer close to the rim of the disk, where the fields are localized.

\begin{figure}[ht]
	\includegraphics[width=\linewidth]{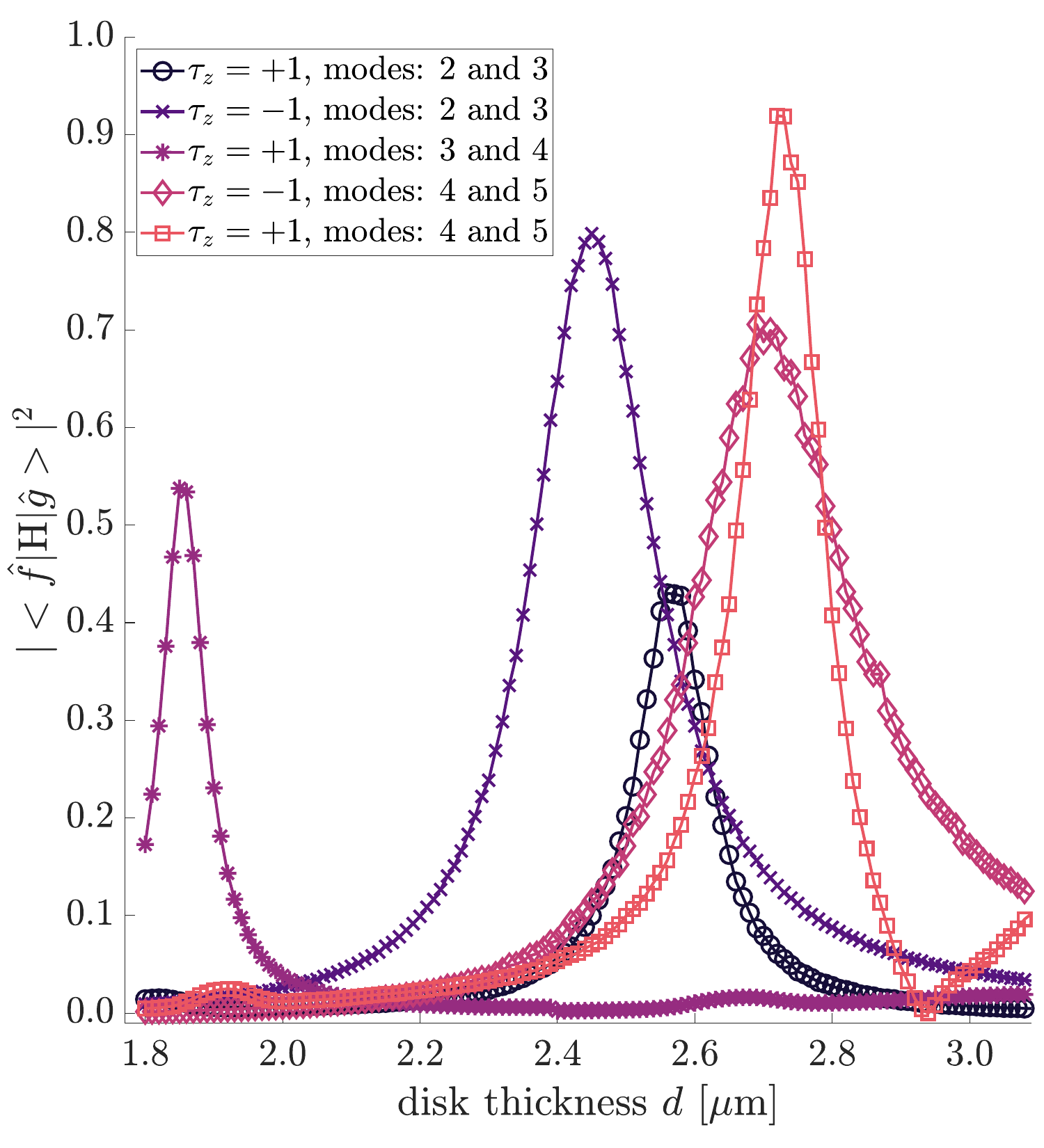}
	\caption{\label{fig:scalar_product} Result of \Eq{eq:normalized_scalar_product} for chosen pairs of modes as a function of disk thickness. The peaks coincide with the thickness for which the anti-crossing occurs between the particular modes, as marked in Fig.~\ref{fig:anticrossings}.}
\end{figure}

\begin{figure}[ht]
	\includegraphics[width=\linewidth,trim=15 5 23 5, clip]{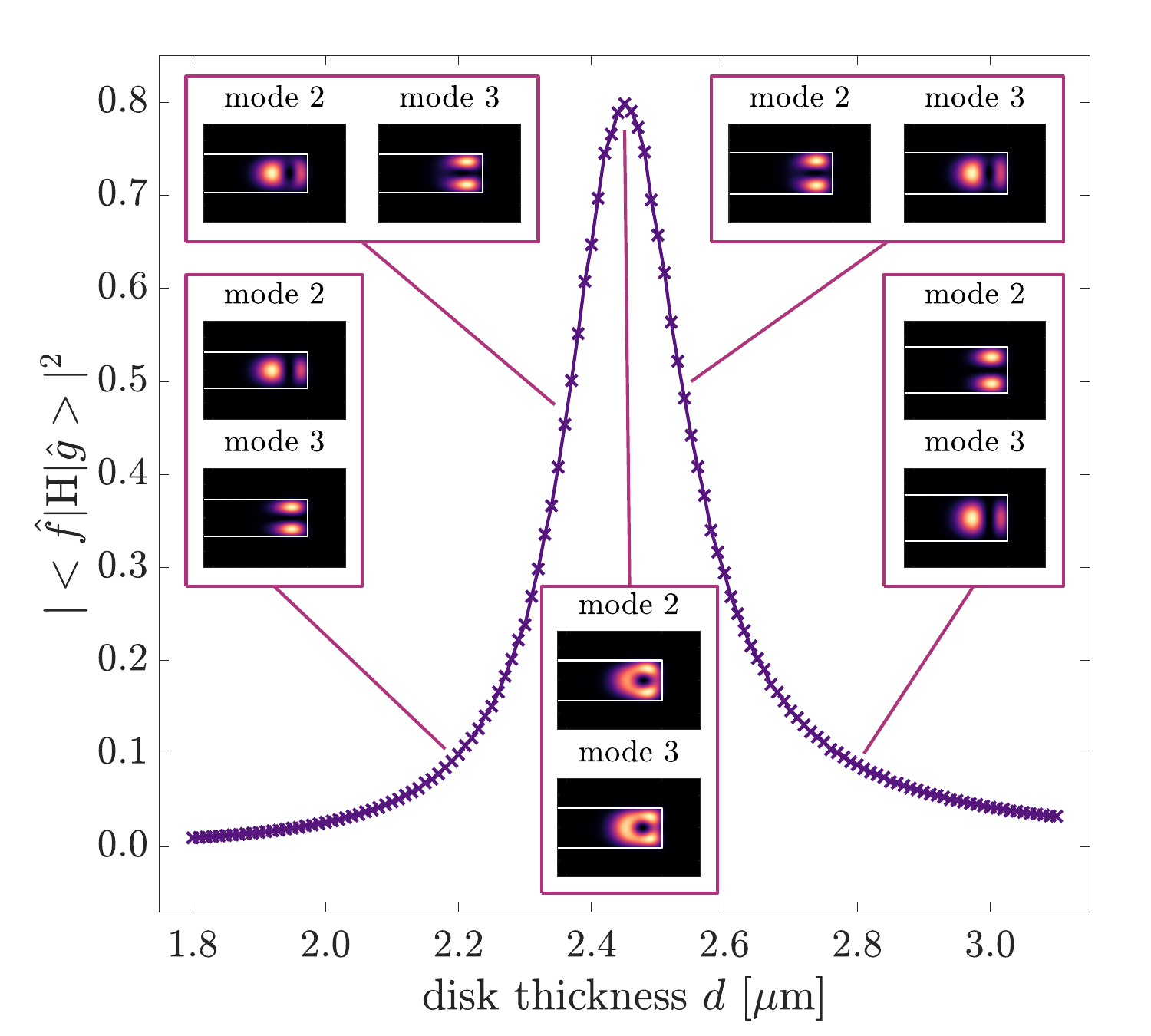}\caption{\label{fig:anticrossing_energy_density} Result of \Eq{eq:normalized_scalar_product} for the modes of spectral order 2 and 3 and $\tau_z=-1$ as a function of the disk thickness. The inset figures show the normalized energy density distribution of the modal fields at particular values of the thickness. The field distributions plotted here are closely related to the ones in \cite[Fig.~3]{woska2022intrinsic}.}
\end{figure}

The modes that cross, having opposite $\tau_z$, are orthogonal under the cross-energy scalar product. This is readily seen by splitting the surface integrals in \Eq{eq:esp} into their $z>0$ and $z<0$ pieces, whose sum (including the sum over $\lambda$) cancels out if the modes have opposite $\tau_z$. Figure~\ref{fig:scalar_product} shows the result of \Eq{eq:normalized_scalar_product} for selected pairs of modes of the same $\tau_z$ and same parameter $\tau_y=1$ as introduced in \Eq{eq:even-odd}. As explained in Appendix~\ref{appendixB}, the values of the scalar products for this example are virtually the same for both signs of $\tau_y$, because the differences are of the order of the inverse of the Q-factors of the modes. In the general case, the results for both values of $\tau_y=\pm1$ should be considered. Each line of Fig.~\ref{fig:scalar_product} shows how a prominent peak grows as the real frequencies of that particular pair of modes get close to each other in Fig.~\ref{fig:anticrossings}. The position of the peak neatly aligns with the thicknesses at which the two modes anti-cross, which here translates into enlarged non-orthogonality of the respective modal radiations. The origin of this non-orthogonality is further elucidated by the example singled out in Fig.~\ref{fig:anticrossing_energy_density}, which shows the value of \Eq{eq:normalized_scalar_product} for a pair of modes of $\tau_z=-1$. The inset figures present the electric field density profiles of the investigated modes for chosen values of disk thickness. As the disk thickness increases, the profiles deform, reach approximately the same shape, and then separate but with interchanged field profiles and modal numbers $(2,3) \leftrightarrow (3,2)$. The cross-energy scalar product peaks exactly when the two mode profiles have maximum overlap. The cross-energy scalar product can be seen as a measure of the similarity between the radiation fields of two modes, in particular in the far-field. It is therefore reassuring that the radiations are most similar exactly when the modal eigenfields inside the resonator are most similar.

\section{Conclusion and Outlook \label{sec:con}}
We have put forward a scalar product for the study of optical resonators using the conformally-invariant scalar product for radiation fields. We showed how any given resonant mode, obtained for example from a Maxwell solver, can be used to determine fields on the surface of the resonator that produce a finite-energy polychromatic emission free of divergences. Then, we identified a cross-energy expression between the radiations of any two given modes as a suitable scalar product. The application to the modes of a disk-shaped WGM resonator showed that the cross-energy scalar product produces physically relevant information, as it predicts the crossing or anti-crossing between modes upon smooth changes of the geometrical parameters of the disk, and provides information regarding the ability to distinguish between the radiation patterns of different leaky modes.

This work opens a path for a different elucidation of the orthogonality and degree of asymptotic completeness of a given series of resonant modes. The same cross-energy scalar product can be used between the radiation of a given mode and any general radiation field, thereby allowing the decomposition of the latter into normalized modes. 

\section*{Acknowledgements}
Maria Paszkiewicz-Idzik acknowledges support by the NHR@KIT program. Lukas Rebholz acknowledges support by the Karlsruhe School of Optics \& Photonics (KSOP). This work was partially funded by the Deutsche Forschungsgemeinschaft (DFG, German Research Foundation) -- Project-ID 258734477 -- SFB 1173. Ivan Fernandez-Corbaton warmly thanks Philippe Lalanne and the members of his group, particularly Tong Wu, for the hospitality and illuminating discussions regarding electromagnetic eigenmodes during an academic visit in the spring of 2024. Ivan Fernandez-Corbaton and Carsten Rockstuhl acknowledge funding by the Helmholtz Association via the Helmholtz program ``Materials Systems Engineering'' (MSE). We are grateful to the company JCMwave for their free provision of the FEM Maxwell solver JCMsuite.
\appendix
\section{A general simplification}\label{appendixA}
  The integrals over $|\mathbf k|$ in \Eq{eq:espA} can be solved analytically, using
\begin{equation}\label{eq:k-integral}
    \int_{0}^\infty \mathrm{d} x \frac{1}{x - z_1^\ast}\,\frac{1}{x - z_2}
    =
    \frac{\Log(-z_1^\ast) - \Log(-z_2)}{-z_1^\ast + z_2}
\end{equation}
for $z_{1,2} \notin \mathbb{R}$. Identifying $x$ with $c_0 |\mathbf k|$ and $(z_1, z_2)$ with one of the combinations $(\omega_f, \omega_g)$, $(\omega_f, -\omega_g^\ast)$, $(-\omega_f^\ast, \omega_g)$, and $(-\omega_f^\ast, -\omega_g^\ast)$, and using \Eq{eq:k-integral} four times yields \Eq{eq:esp}.

\section{Application to an achiral disk}\label{appendixB}
In the case of cylindrical symmetry, the surface integrals in \Eq{eq:esp} can be rewritten using the azimuthal dependence of the modes described in \Eq{eq:M_JCM}, for example from the first line of \Eq{eq:esp}:
\begin{multline}\label{eq:surface_integral}
    \int_{\rr\in\partial D}\mathrm d\mathbf S(\rr)\cdot\left[ 
        \left(\mathbf{M}_{\lambda,f}^{(m^f)}(\rr)\right)^\ast \times \mathbf{M}_{\lambda,g}^{(m^g)}(\rr)
    \right]
    \\
    =
    \int_{\mathcal{C}}\mathrm ds \rho \hat{\mathbf v}(\rr)\cdot\left[ 
        \left(\mathbf{M}_{\lambda,f}^{(m^f)}(\rr)\right)^\ast \times \mathbf{M}_{\lambda,g}^{(m^g)}(\rr)
    \right]
    \\\times\int_0^{2\pi}\mathrm d\varphi\mathrm{e}^{\ii (-m^f + m^g) \varphi} \, .
\end{multline}
Here, we have written $\mathrm d \mathbf S(\rr)=\mathrm d s \cdot \rho \mathrm d\varphi \hat{\mathbf v}(\rr)$ in cylindrical coordinates $(\rho, \varphi, z)$, with the surface normal vector $\hat{\mathbf v}$. The contour $\mathcal C$ refers to the 1D curve defined as the $\varphi=0$ slice of $\partial D$, and $\mathrm d s$ denotes a differential line element of $\mathcal C$. The integral over $\varphi$ in \Eq{eq:surface_integral} evaluates to $2\pi$ only in the case where $m^f = m^g$, making the entire surface integral vanish otherwise.

Analogously, we can simplify the surface integral in the second line of \Eq{eq:esp}, which vanishes unless ${m^f = -m^g}$. Note that the surface integrals in lines three and four of \Eq{eq:esp} are simply the complex conjugates of the integrals in lines two and one, respectively.

In summary, only pairs of modes satisfying ${\left|m^f\right| = \left|m^g\right|}$ can have a nonzero scalar product. In either case (excluding $m = 0$), the four terms in \Eq{eq:esp} are reduced to only two, with the two-dimensional surface integrals
\begin{equation*}
    \int_{\rr\in\partial D}\mathrm d\mathbf S(\rr)\cdot\big[\ldots\big]
\end{equation*}
replaced by one-dimensional line integrals
\begin{equation*}
    2\pi\int_{\mathcal{C}}\mathrm ds \rho \hat{\mathbf v}(\rr)\cdot\big[\ldots\big]\,.
\end{equation*}

As explained in the main text, we can build mirror symmetric modes [\Eq{eq:even-odd}] by combining $\mathbf{M}_{\lambda,f}^{(m^f)}(\rr)$ fields with their mirror reflections $\mathbf{M}_{\lambda,f}^{(-m^f)}(\rr)$. In a cylindrically symmetric system that also feature a plane of mirror symmetry containing the axis of rotation, $\mathbf{M}_{\lambda,f}^{(-m^f)}(\rr)$ can be obtained from $\mathbf{M}_{\lambda,f}^{(m^f)}(\rr)$ using one of the mirror symmetries of the system where the axis of rotational symmetry is contained in the mirror plane. For example, using ${\op{M}_y=\Big[\begin{smallmatrix}1&0&0\\0&-1&0\\0&0&1\end{smallmatrix}\Big]}$ in Cartesian coordinates, which maps ${y \mapsto -y}$, we can write
\begin{equation}\label{eq:explicit-mirror}
	\mathbf{M}_{\lambda,f}^{(-m^f)}(\rr) = \op{M}_y \mathbf{M}_{-\lambda,f}^{(m^f)}(\op{M}_y \rr) \,.
\end{equation}
Note that \Eq{eq:explicit-mirror} relates fields of opposite helicity, in line with the fact that any mirror transformation changes an eigenstate of helicity into one of opposite helicity, or, colloquially, flips its handedness. This change of handedness can also be deduced from the transformations of the modal fields under a mirror transformation, ${\ModeEf \mapsto \op{M}_y\boldsymbol{\mathcal{E}}_f(\op{M}_y\rr)}$ and ${\ModeHf \mapsto -\op{M}_y\boldsymbol{\mathcal{H}}_f(\op{M}_y\rr)}$, whose difference in sign is due to the polar and axial character of electric and magnetic fields, respectively.

Above arguments can be used to show the mutual orthogonality of the even and odd modal fields introduced in \Eq{eq:even-odd}. For the purpose of brevity, we introduce the notation
\begin{align}
    \ket{f,\pm m^f} &\equiv \Big\{\mathbf{M}_{\lambda,f}^{(\pm m^f)}(\rr),\omega_f\Big\}\\
    \ket{f,\tau_y^f} &\equiv \Big\{\Mlambdafrtau,\omega_f\Big\} \,,
\end{align}
where the fields on the right-hand side correspond to the ones in \Eq{eq:even-odd}, and the meaning of the equivalence sign is the radiation field from the leaky modes as built with \Eq{eq:k}.

Considering the mirror transformation used to obtain the $\mathbf{M}_{\lambda,f}^{(-m^f)}(\rr)$ from $\mathbf{M}_{\lambda,f}^{(m^f)}(\rr)$, one can show that
\begin{equation}
	\begin{split}
		\braket{f,m^f|\mathrm H|g,{-m^g}}& = \braket{f,{-m^f}|\mathrm H|g,{m^g}}\text{, and}\\
		\braket{f,m^f|\mathrm H|g,{m^g}}& = \braket{f,{-m^f}|\mathrm H|g,{-m^g}} \,,
	\end{split}
\end{equation}
and with that
\begin{align}\label{eq:orthogonality-tau}
    \braket{f,\tau_y^f|\mathrm H|g,\tau_y^g} 
    &= \frac{1+\tau_y^f\tau_y^g}{2}\braket{f,{m^f}|\mathrm H|g,{m^g}} \nonumber\\
    &\quad{}+ \frac{\tau_y^f + \tau_y^g}{2} \braket{f,{-m^f}|\mathrm H|g,{m^g}} \nonumber\\
    &= \Big(\braket{f,{m^f}|\mathrm H|g,{m^g}} \nonumber\\
    &\quad{}+ \tau_y^f\braket{f,{-m^f}|\mathrm H|g,{m^g}}\Big)\delta_{\tau_y^f,\tau_y^g} \,,
\end{align}
where the Kronecker delta in the last line formalizes the orthogonality of modes with nonmatching $\tau_y$.

The remaining $\tau_y^f$ in the last line of \Eq{eq:orthogonality-tau} shows that $\braket{f,\tau_y^f=1|\mathrm H|g,\tau_y^g=1}$ and $\braket{f,\tau_y^f={-1}|\mathrm H|g,\tau_y^g={-1}}$ are in general different. Such difference is on the order of the inverse Q-factor of the modes. This can be seen from the pre-factors involving the modal frequencies in \Eq{eq:esp}. For ${m^f = m^g}$, the term ${\braket{f,{m^f}|\mathrm H|g,{m^g}}}$, involves evaluating the first and fourth lines in \Eq{eq:esp}, and the term $\braket{f,{-m^f}|\mathrm H|g,{m^g}}$, involves lines two and three in \Eq{eq:esp}. For degenerate modes, but also for spectrally-close modes, the absolute value of the ratio between the pre-factors of such lines is roughly the Q-factor of the modes. The relative difference between $\braket{f,\tau_y|\mathrm H|g,\tau_y}$ for different $\tau_y$ is correspondingly small for the modes of the disk that we study.

\bibliography{main}

\end{document}